\documentclass{article} 
\usepackage{epsfig} 
\usepackage{cite} 
\usepackage[intlimits,reqno]{amsmath}
\def\eqref#1{Eq.(\ref{#1})}

\def\itref#1{(\ref{#1})}

\def\ep{\epsilon}
\def\oo{\infty}
\def\ln{\hbox{ln}} 
 
\def\ie{{\it i.e.}\phantom{.}}

\begin{document} 
\unitlength1cm 
\begin{titlepage} 
\vspace*{-1cm} 
\begin{flushright} 
CERN-TH/2003-277\\ 
TTP03-36\\ 
November 2003 
\end{flushright} 
\vskip 3.5cm 
\renewcommand{\topfraction}{0.9} 
\renewcommand{\textfraction}{0.0} 

\begin{center} 
\boldmath 
{\Large\bf The analytic value of a 4-loop sunrise graph 
in a particular kinematical configuration. 
}\unboldmath 
\vskip 1.cm 
{\large S. Laporta,}$^{a,b,}$ 
\footnote{{\tt e-mail: Stefano.Laporta@bo.infn.it}}
{\large P. Mastrolia\,}$^{c,d,}$ 
\footnote{{\tt e-mail: Pierpaolo.Mastrolia@bo.infn.it}} and 
{\large E. Remiddi\,}$^{e,c,a,}$ 
\footnote{{\tt e-mail: Ettore.Remiddi@bo.infn.it}} 
\vskip .7cm 
\vskip .4cm 
{\it 
  $^a$ INFN, Sezione di Bologna, I-40126 Bologna, Italy \\ 
  $^b$ Dipartimento di Fisica, Universit\`{a} di Parma, I-43100 Parma, 
       Italy \\ 
  $^c$ Dipartimento di Fisica, Universit\`{a} di Bologna, I-40126 Bologna, 
       Italy \\
  $^d$ Institut f\"ur Theoretische Teilchenphysik,
       Universit\"at Karlsruhe, D-76128 Karlsruhe, Germany \\ 
  $^e$ Theory Division, CERN, CH-1211 Geneva 23, Switzerland \\ 
} 
\end{center} 
\vskip 2.6cm 
\begin{abstract} 
The 4-loop sunrise graph with two massless lines, two lines of equal 
mass $M$ and a line of mass $m$, for external invariant timelike and 
equal to $m^2$ is considered. We write differential equations 
in $x=m/M$ for the Master Integrals of the problem, which we Laurent-expand 
in the regularizing continuous dimension $d$ around $d=4$, and then 
solve exactly in $x$ up to order $(d-4)^3$ included; 
the result is expressed in terms of Harmonic PolyLogarithms of 
argument $x$ and maximum weight 7. As a by product, 
we obtain the $x=1$ value, expected to be relevant in QED 4-loop 
static quantities like the electron $(g-2)$. The analytic results 
were checked by an independent precise numerical calculation. 

\end{abstract} 
\vfill
\end{titlepage}

\newpage 
\renewcommand{\theequation}{\mbox{\arabic{section}.\arabic{equation}}} 
\section{Introduction} 
\label{sec:int} 
\begin{figure}[h] 
\begin{center} 
\includegraphics*[2cm,1cm][14cm,6cm]{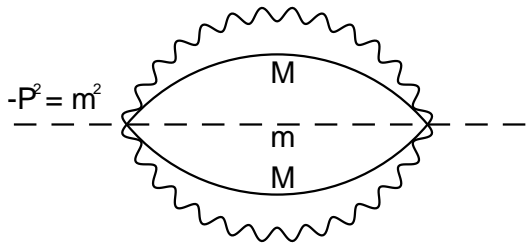} 
\caption{The considered 4-loop sunrise graph.} 
\end{center}
\label{fig:1} 
\end{figure}

\setcounter{equation}{0} 
This paper is devoted to the analytic evaluation of the Master Integrals 
(MI's) associated to the 4-loop sunrise graph with two massless lines, 
two massive lines of equal mass $M$, 
another massive line of mass $m$, with $m \ne M$, and the external invariant 
timelike and equal to $m^2$, as depicted in Fig.~\ref{fig:1}. \par 
We will follow for the analytic integration 
the differential equation method already proposed 
in~\cite{Kotikov:1990kg}, further developped in~\cite{Remiddi:1997ny} 
(and then used in~\cite{Caffo:1998du},\cite{Gehrmann:1999as} and in 
many subsequent applications which would be too long to report here), 
as well as the finite difference method~\cite{lap2},\cite{lap} for 
an independent numerical check. \par 
The differential equation method was already followed in two similar two- 
and three-loop calculations~\cite{sun2},~\cite{sun3}; the fact that its 
use could be extended without major changes to the present four-loop 
calculation witnesses for its generality and power. Among the 
advantages of the method, it allows a rather clear separation of 
the merely algebraic part of the work (which is, not surprisingly, 
always very heavy in this kind of multiloop calculations, and can be 
most conveniently processed by a computer algebra program, in our case 
{\tt FORM}~\cite{FORM}), from the really analytic issues of the problem, 
which can then be better investigated without the disturbance of the 
algebraic complexity. In our case, indeed, the heart of the analytic 
calculation was the study of a homogeneous fourth order differential 
equation, whose solutions turned out to be, in a remarkably simple way, 
either a rational fraction 
or repeated quadratures of rational fractions. The required four-loop 
integral could then be obtained almost mechanically by repeated 
quadratures in terms of Harmonic PolyLogarithms~\cite{hpl}. \par 
Several other different approaches to the analytic evaluation of 
multiloop integrals are available in the literature, such as the powerful 
asymptotic expansion method (a fairly complete account can be found in the 
recent book~\cite{Smirnov}), and it would be interesting to compare 
the advantages and drawbacks of the various methods for the exacting, 
four-loop integration which we consider; but as the results of the 
present paper are new, a meaningful comparison cannot yet be carried out. 
\par 
Following, as already said above, the approach already used in ~\cite{sun2} 
and ~\cite{sun3}, we identify the MI's of the current 
problem within the continuous 
$d$-dimensional regularization, write the system of differential equations 
in $x=m/M$ satisfied by the MI's, convert it into a higher order 
differential equation for a single MI, Laurent-expand in $(d-4)$ around 
$d=4$, solve the associated homogeneous 
equation at $d=4$ (as in previous cases, the solutions of the homogeneous 
equation are surprisingly simple) and then use recursively Euler's method of 
the variation of the constants for obtaining the coefficient of the 
$(d-4)$ expansion in closed analytic form. The result involves 
Harmonic PolyLogarithms (HPL's)~\cite{hpl} of argument $x$ and weight 
increasing with the order in $(d-4)$. We push the analytic calculation, 
which works up to virtually any order in $(d-4)$, up to $(d-4)^{3}$ 
included, involving HPL's of weight up to $w=7$ 
included. The integration constants are fixed at 
$x=0\ $; as a by product we obtain the values at $x=1$, which are 
relevant in the evaluation of 4-loop static quantities such as the 
electron $(g-2)$ in QED. The result was checked and confirmed by an 
independent numerical calculation performed with the method 
of~\cite{lap2},~\cite{lap}. 
\par 
The plan of the paper is the following. In Section~\ref{sec:MI} we define 
the Master Integrals and write the differential equations; in 
Section~\ref{sec:xto0p} 
we study the $x\to0$ behaviour; in Section~\ref{sec:expd-4} we work out the 
Laurent-expansion in $(d-4)$ and discuss the associated homogeneous 
equation; in Section~\ref{sec:solved-4} we write down the solutions by 
Euler's method of the variation of the constants and give the $x=1$ values; 
Section~\ref{sec:numcal} deals with the independent numerical calculation 
by which we check the results at $x=1$ of the previous Section. 
in Section~\ref{sec:xto0val} we carry out the evaluation of the $x\to0$ 
values by direct integration in the parametric space. 

\section{The Master Integrals and the Differential Equations} 
\label{sec:MI} 
\setcounter{equation}{0} 
We find that the problem has 5 MI's, which we choose to be 
\begin{equation} 
  F_i(d,M^2,m^2,P^2=-m^2) = \frac{1}{(2\pi)^{4(d-2)}} 
     \int \frac{d^dk_1\  d^dk_2 \ d^dk_3 \ d^dk_4 \ N_i} 
      { k_1^2 k_2^2 (k_3^2+M^2) (k_4^2+M^2) 
                                [(P-k_1-k_2-k_3-k_4)^2+m^2] } \ , 
\label{eq:defFi} 
\end{equation} 
where the 5 numerators $N_i$ are 
$(M^2, k_1 \cdot k_3, p \cdot k_3, k_1 \cdot k_2, p \cdot k_2)$. 
In terms of the dimensionless variable $ x = m/M $ and putting 
$P=Mp$ one can introduce 5 dimensionless functions $\Phi_i(d,x)$ through 
\begin{equation} 
     F_i(d,M^2,m^2,P^2=-m^2) = M^{4d-8} \ C^4(d)\ \Phi_i(d,x) \ , 
\label{eq:FtoPhi} 
\end{equation} 
where $C(d)= (4\pi)^\frac{4-d}{2}\Gamma(3-d/2) $ is an overall loop 
normalization factor, with the limiting value $C(4)=1$ at $d=4$. 
Some of the formulae which will follow (in particular the 
differential equations) are slightly simpler when written in terms of 
$x^2$ rather than $x$; but as $x$ is the most convenient variable 
for expressing the final analytic results, we stick to $x$ from 
the very beginning. 
\par 
As in~\cite{sun2}\cite{sun3}, 
the derivation of the system of differential equations is 
straightforward; the derivatives of the MI's, \ie of the 5 functions 
$\Phi_i(d,x)$, with respect to $x$ are easily 
carried out in their representation as loop-integrals Eq.(\ref{eq:defFi}); 
when the result is in turn expressed in terms of the same MI's, one 
obtains the following linear system of first order differential equations 
in $x$ 
\begin{eqnarray}
\frac{d \Phi_1(d,x)}{d x}
 & = &
        \Bigg\{ 
                 \frac{(3d - 7)}{x} 
               + \frac{3 \ (d-2)}{2 \ (1-x)}
               - \frac{3 \ (d-2)}{2 \ (1+x)}
        \Bigg\} \Phi_1(d,x) 
      - \Bigg\{ 
                 \frac{3(d - 2)}{x} 
               + \frac{3 \ (d-2)}{2 \ (1-x)}
        \nonumber \\
        & &
               - \frac{3 \ (d-2)}{2 \ (1+x)}
        \Bigg\}
        \bigg(
                 3 \Phi_2(d,x) 
               - 3 \Phi_3(d,x) 
               +   \Phi_4(d,x)
               - 3 \Phi_5(d,x)
        \bigg) \ ,
\label{1stDEPhi1} \\
 & & \nonumber \\
\frac{d \Phi_2(d,x)}{d x}
 & = & - \frac{(d-2)}{x} \bigg(
                                  \Phi_2(d,x)
                                - 2 \Phi_3(d,x)
                                + \Phi_4(d,x)
                                - 2 \Phi_5(d,x)
                         \bigg) \ ,
\label{1stDEPhi2} \\
 & & \nonumber \\
\frac{d \Phi_3(d,x)}{d x}
& = &
       - \Bigg\{ 
                 \frac{3 \ (d-2)}{2 \ (1-x)}
               - \frac{3 \ (d-2)}{2 \ (1+x)}
         \Bigg\} \bigg(
                         \Phi_1(d,x)
                       - 3 \Phi_2(d,x)
                       - \Phi_4(d,x)
                 \bigg) 
         \nonumber \\
         & &
       - \Bigg\{ 
                 \frac{3(d - 2)}{x} 
               + \frac{9 \ (d-2)}{2 \ (1-x)}
               - \frac{9 \ (d-2)}{2 \ (1+x)}
         \Bigg\} \bigg(
                         \Phi_3(d,x)
                       + \Phi_5(d,x)
                 \bigg) \ ,
\label{1stDEPhi3} \\
 & & \nonumber \\
\frac{d \Phi_4(d,x)}{d x}
 & = & \frac{2(d-2)}{x} \bigg(
                                  \Phi_2(d,x)
                                + \Phi_4(d,x)
                         \bigg) \ ,
\label{1stDEPhi4} \\
 & & \nonumber \\
\frac{d \Phi_5(d,x)}{d x}
 & = & \frac{2(d-2)}{x} \bigg(
                                  \Phi_3(d,x)
                                + \Phi_5(d,x)
                         \bigg) \ ,
\label{1stDEPhi5}
\end{eqnarray}

At variance with the cases discussed in~\cite{sun2},\cite{sun3}, 
the system is homogeneous; indeed, quite in general the non homo\-ge\-neous 
terms are given by the MI's of the ``subtopologies" of the considered graph, 
obtained by shrinking to a point any of its propagator lines. When that 
is done for the 5-propagator ``topology" of Fig.(\ref{fig:1}), 
one obtains the product of 4 tadpoles; but as the considered graph has 
two massless propagators, at least one massless tadpole is always 
present in the product; as in the $d$-dimensional regularization 
massless tadpoles vanish, the product of the 4 tadpoles is always equal 
to zero -- and therefore the differential equations are homogeneous. 
\par 
By inspection, one sees that $\Phi_3(d,x), \Phi_5(d,x)$ appear in the 
r.h.s. of Eq.s(\ref{1stDEPhi1}-\ref{1stDEPhi5}) only in the combination 
\begin{equation} 
 \Psi_3(d,x) = \Phi_3(d,x) + \Phi_5(d,x) \ ; 
\label{defPsi3} 
\end{equation} 
the other linearly independent combination of the two MIs, say 
\begin{equation} 
 \Psi_5(d,x) = \Phi_3(d,x) - \Phi_5(d,x) \ , 
\label{defPsi5} 
\end{equation} 
decouples and can be expressed in terms of the other integrals by means 
of the trivial 1st order differential equation 
\begin{eqnarray} 
\frac{d \Psi_5(d,x)}{d x} 
 & = & 
       - \Bigg\{ 
                 \frac{3 \ (d-2)}{2 \ (1-x)} 
               - \frac{3 \ (d-2)}{2 \ (1+x)} 
         \Bigg\} \bigg( 
                         \Phi_1(d,x) 
                       - 3 \Phi_2(d,x) 
                       - \Phi_4(d,x) 
                 \bigg) 
         \nonumber \\ 
         & & 
       - \Bigg\{ 
            \frac{5 (d-2)}{x} 
          + \frac{9 \ (d-2)}{2 \ (1-x)} 
          - \frac{9 \ (d-2)}{2 \ (1+x)} 
         \Bigg\} \Psi_3(d,x) 
\label{Psi5eq} 
\end{eqnarray} 
As $\Psi_5(d,x)$ does not enter in the r.h.s. of 
Eq.s(\ref{1stDEPhi1}-\ref{1stDEPhi5}) the 4 linear 
equations for $\Phi_1(d,x),\ \Phi_2(d,x),\ \Psi_3(d,x)$ and 
$\Phi_4(d,x)$ can be written as a fourth order equation for the first 
Master Integral $\Phi_1(x)$, which will be called simply $\Phi(d,x)$ from 
now on, and which is therefore equal to 
\begin{equation} 
  \Phi(d,x) = \frac{C^{-4}(d)}{(2\pi)^{4(d-2)}} \int 
     \frac{d^dk_1\  d^dk_2 \ d^dk_3 \ d^dk_4 } 
      { k_1^2 k_2^2 (k_3^2+1) (k_4^2+1) 
        [(p-k_1-k_2-k_3-k_4)^2+x^2] } \ , \ (p^2=-x^2) \ . 
\label{eq:defPhi} 
\end{equation} 
One obtains for $\Phi(d,x)$ the following 4-th order differential equation 
\begin{eqnarray} 
  x^3 (1-x^2) \frac{d^4 \Phi(d,x)}{d x^4}
+ x^2 \bigg\{ 1 + 5 x^2 - 3 (d-4) (1 - 3 x^2)
      \bigg\} \frac{d^3 \Phi(d,x)}{d x^3}
&& \nonumber\\ 
- x \bigg\{
            12
          + 6 x^2
          + (d-4) (13 + 32 x^2)
          + (d-4)^2 (1 + 26 x^2)
    \bigg\} \frac{d^2 \Phi(d,x)}{d x^2} 
&& \nonumber\\ 
+  \bigg\{
            12
          - 18 x^2
          + (d-4) (25 - 2 x^2)
          + 8 (d-4)^2 (2 + 5 x^2)
          + 3 (d-4)^3 (1 + 8 x^2)
   \bigg\} \frac{d \Phi(d,x)}{d x}
&& \nonumber\\ 
+  4 x \bigg\{ 
          + 12 
          + 29 (d-4)
          + 23 (d-4)^2
          + 6 (d-4)^3
     \bigg\} \Phi(d,x) 
&& {\kern-15pt} =\ 0 \ . 
\label{eq:4thordeq} 
\end{eqnarray} 
\par 
The expression of the other MI's in terms of $\Phi(d,x)$ and its first 
3 $x$-derivatives reads 
\begin{eqnarray}
\Phi_2(d,x) &=&
        \bigg\{
            \frac{1}{2}
          - \frac{1}{6 \ (d-2)}
          + \frac{x^2}{5 \ (d-2)^3} 
          - \frac{2 x^2}{3 \ (d-2)^2} 
          + \frac{7 x^2}{15 \ (d-2)} 
        \bigg\} \Phi(d,x) \nonumber \\
 & &
       - \bigg\{
            \frac{x (1 - 7 x^2)}{10 \ (d-2)^3} 
          + \frac{x (1 + 16 x^2)}{12 \ (d-2)^2} 
          - \frac{x (11 + 28 x^2)}{60 \ (d-2)} 
        \bigg\} \frac{d \Phi(d,x)}{d x} \nonumber \\
 & &
       - \bigg\{ 
            \frac{x^2 (1 - 2 x^2)}{5 \ (d-2)^3} 
          - \frac{x^2 (1 - 10 x^2)}{30 \ (d-2)^2} 
         \bigg\} \frac{d^2 \Phi(d,x)}{d x^2} \nonumber \\
 & &
       - \frac{x^3 (1 - x^2)}{20 \ (d-2)^3} 
            \frac{d^3 \Phi(d,x)}{d x^3} \ ,  \label{Phi2fromPhi} \\
 & & \nonumber \\
    \Psi_3(d,x) &=&
       - \bigg\{
            \frac{x^2}{15 \ (d-2)^3} 
          - \frac{1 x^2}{3 \ (d-2)^2} 
          + \frac{3 x^2}{5 \ (d-2)} 
        \bigg\} \Phi(d,x) \nonumber \\
 & &
       + \bigg\{
            \frac{x (1 - 7 x^2)}{30 \ (d-2)^3} 
          - \frac{x (1 - 8 x^2)}{12 \ (d-2)^2} 
          + \frac{x (1 - 12 x^2)}{20 \ (d-2)} 
        \bigg\} \frac{d \Phi(d,x)}{d x} \nonumber \\
 & &
       + \bigg\{ 
            \frac{x^2 (1 - 2 x^2)}{15 \ (d-2)^3} 
          - \frac{x^2 (2 - 5 x^2)}{30 \ (d-2)^2} 
         \bigg\} \frac{d^2 \Phi(d,x)}{d x^2} \nonumber \\
 & &
       - \frac{x^3 (1 - x^2)}{60 \ (d-2)^3} 
            \frac{d^3 \Phi(d,x)}{d x^3} \ , \label{Psi3fromPhi} \\
 & & \nonumber \\
\Phi_4(d,x) &=&
       - \bigg\{
            \frac{1}{2}
          - \frac{1}{6 \ (d-2)}
          + \frac{4x^2}{5 \ (d-2)^3} 
          - \frac{3 x^2}{ (d-2)^2} 
          + \frac{43 x^2}{15 \ (d-2)} 
        \bigg\} \Phi(d,x) \nonumber \\
 & &
       + \bigg\{
            \frac{2 x}{5 \ (d-2)^3} 
          - \frac{x (14 - 30 x^2)}{5 \ (d-2)^2} 
          - \frac{x (11 + 43 x^2)}{15 \ (d-2)} 
        \bigg\} \frac{d \Phi(d,x)}{d x} \nonumber \\
 & &
       + \bigg\{ 
            \frac{4x^2 (1 - 2 x^2)}{5 \ (d-2)^3} 
          - \frac{3x^2 (1 - 5 x^2)}{10 \ (d-2)^2} 
         \bigg\} \frac{d^2 \Phi(d,x)}{d x^2} \nonumber \\
 & &
       + \frac{x^3 (1 - x^2)}{5 \ (d-2)^3} 
            \frac{d^3 \Phi(d,x)}{d x^3} \ . \label{Phi4fromPhi} 
\end{eqnarray}

\section{The $x\to 0$ behaviour of $\Phi(d,x)$ } 

\label{sec:xto0p} 
\setcounter{equation}{0} 
By inspection, one finds that the most general solution of 
Eq.(\ref{eq:4thordeq}) can be expanded for $x\to 0$ in the form 
\begin{equation} 
  \Phi(d,x) = \sum_{i=1}^4 x^{\alpha_i} 
              \left( \sum_{n=0}^\infty A_n^{(i)}(d) x^{2n} \right) \ , 
\label{eq:xexpPhi} 
\end{equation} 
where the values of the 4 exponent $\alpha_i$ are 
\begin{eqnarray} 
   \alpha_1 &=& 0 \ , \nonumber\\ 
   \alpha_2 &=& (d-2) \ , \nonumber\\ 
   \alpha_3 &=& -(d-2) \ , \nonumber\\ 
   \alpha_4 &=& (3d-7) \ ; 
\label{eq:alphaval} 
\end{eqnarray} 
the $A_0^{(i)}(d)$ are the 4 integration constants, and all the other 
coefficients $A_n^{(i)}(d)$ for $n>0$ are determined by the differential 
equation Eq.(\ref{eq:4thordeq}) once the integration constants are fixed. 
\par 
It is interesting to recall the leading exponents of the $x\to0$ 
expansions of the 
solutions of the corresponding equations encountered in~\cite{sun2} 
and~\cite{sun3}. Calling $\alpha_i^{(2)}$ the exponents for the 
2-loop graph of~\cite{sun2} and $\alpha_i^{(3)}$ those of the 
3-loop graph of~\cite{sun3}, one finds that there are always 4 exponents. 
The explicit values in the case of~\cite{sun2} are 
\begin{eqnarray} 
   \bullet\ \ \ \alpha_1^{(2)} &=& 0 \ , \nonumber\\ 
   \bullet\ \ \ \alpha_2^{(2)} &=& (d-2) \ , \nonumber\\ 
   \alpha_3^{(2)} &=& -(d-2) \ , \nonumber\\ 
   \alpha_4^{(2)} &=& (d-3) 
\label{eq:alpha2val} 
\end{eqnarray} 
and in the case of~\cite{sun3} 
\begin{eqnarray} 
   \alpha_1^{(3)} &=& 0 \ , \nonumber\\ 
   \bullet\ \ \ \alpha_2^{(3)} &=& (d-2) \ , \nonumber\\ 
   \alpha_3^{(3)} &=& -(d-2) \ , \nonumber\\ 
   \alpha_4^{(3)} &=& (2d-5) \ , 
\label{eq:alpha3val} 
\end{eqnarray} 
where the exponents marked by a bullet ($\bullet$) correspond to the 
behaviours forced by the inhomogeneous terms. 
The similarity between the 3 sets of behaviours is impressive: the 
first 3 exponents are identical, the fourth differ in steps of $(d-2)$ 
for each additional loop. 
\par 
In more details, 2 of the exponents of Eq.s(\ref{eq:alpha2val}) correspond 
to the 2 independent solutions of the associated homogeneous differential 
equation, which is of 2nd order, while the other 2 are forced by the 2 
independent behaviours for $x\to 0$ developed by the 
inhomogeneous terms (both products of 2 tadpoles, the first product 
of two tadpoles of mass $M$, the second of a tadpole of mass $M$ and a 
tadpole of mass $m=Mx$). In the $x\to 0$ expansion of the most general 
solution, one is therefore left with 2 undetermined integration constants, 
corresponding to the two homogeneous solutions, while all the terms 
with the behaviours of the inhomogeneous terms are fully determined by the 
inhomogeneous equation itself. 
Similarly, in Eq.s(\ref{eq:alpha3val}) 3 exponents correspond to the 3 
solutions 
of the 3rd order homogeneous equation, with the 4th exponent
forced by  
the inhomogeneous term (product of 3 tadpoles, of masses $M, M$ and $m=Mx$), 
so that there are in principle 3 undetermined integration constants, 
the term corresponding to the remaining behaviour being fixed by the 
equation. In the present case, finally, the equation is homogeneous 
(as already observed, there 
is no inhomogeneous term, as all the possible products of 4 tadpoles 
involve at least a vanishing zero-mass tadpole) and the 4 exponents of 
Eq.s(\ref{eq:alphaval}) correspond to the behaviour of the 4 homogeneous 
solutions, so that one is left in principle with 4 undetermined 
integration constants. 
\par 
A qualitative inspection of the integrals which one tries to evaluate by 
means of the differential equations (Eq.(\ref{eq:defPhi}) of the present 
paper, Eq.(1.3) of ~\cite{sun2} and Eq.(2.3) of~\cite{sun3}) shows 
that they are all finite (just finite, not analytic!) for $x\to 0^+$ 
and $(d-2)>0$; that is sufficient to rule out from their expression as 
solutions of the differential equation the terms 
with the behaviour of the third and the fourth exponent (which is 
negative when $d$ is just above 2). 
\par 
In the case of~\cite{sun2}, that fixes completely the solution. 
In the case of~\cite{sun3}, one integration constant is left undetermined; 
to fix it, one has to provide some independent information, such as 
the value of the required Feynman integral at $x=0$ (which corresponds 
to a simpler vacuum amplitude); 
that value can be provided by an explicit ``conventional" calculation, 
say in parameter space, which is in any case much easier than a 
calculation for non-zero values of the variable $x$. (A closer analysis 
carried out in~\cite{sun3} shows however that the regularity of the 
integral at $x=1$ is sufficient to fully determine the solution, 
so that the actual knowledge of the $x=0$ value can be used as an 
independent check). 
\par 
In the current case, as the equation for $\Phi(d,x)$ is homogeneous, the 
only information is that $A_0^{(3)}(d)$ and $A_0^{(4)}(d)$ are both equal 
to zero due to the finiteness for $x\to0^+$; by substituting the 
{\it ansatz} Eq.(\ref{eq:xexpPhi}) in Eq.(\ref{eq:4thordeq}) and dropping 
$A_0^{(3)}(d), A_0^{(4)}(d),\ $ one finds for $\Phi(d,x)$ Eq.(\ref{eq:defPhi}) 
the $x\to 0$ expansion 
\begin{eqnarray} 
\Phi(d,x) &=& 
            A_0^{(1)}(d) 
              \left(   1 
                     - \frac{2 (2d-5) (3d-8)}{3 d (d-4)} \ x^2 
                     + O(x^4) 
              \right) \nonumber\\ 
          &+& A_0^{(2)}(d)\ x^{d-2} 
              \left(   1 
                     + \frac{(d-3)(d-4)(3d-8)}{2 d (2d-7)} \ x^2 
              + O(x^4) \right) \ . 
\label{eq:Phiexpexpl} 
\end{eqnarray} 
The expansion depends on the two as yet unspecified integrations 
constants $A_0^{(1)}(d), A_0^{(2)}(d)$ -- which are to be 
provided by an independent, explicit calculation. 
That is done in Section~\ref{sec:xto0val}, see Eq.(\ref{eq:Aval}), by 
direct integration in the parametric space. 

Let us note here, for completeness, that in the present case the 
knowledge of the regularity of the solution at $x=1$ does not provide 
any additional information. 

\section{The expansion in $(d-4)$ and the homogeneous equation 
         at $d=4$.} 
\label{sec:expd-4} 
\setcounter{equation}{0} 
The Laurent's expansion in $(d-4)$ of $\Phi(d,x)$ Eq.(\ref{eq:defPhi}) is 
\begin{equation} 
   \Phi(d,x) = \sum_{n=-4}^\infty (d-4)^n \Phi^{(n)}(x) \ , 
\label{eq:dexpPhi} 
\end{equation} 
as it is known on general grounds that it develops at most a fourth 
order pole in $(d-4)$. By substituting in Eq.(\ref{eq:4thordeq}) 
one obtains a system of inhomogeneous, chained equations for the 
coefficients $\Phi^{(n)}(x)$ of the expansion in $(d-4)$; the 
generic equation reads 
\begin{equation} 
 \left[ 
  x^3 (1-x^2) \frac{d^4}{d x^4}
+ x^2 (1 + 5 x^2) \frac{d^3}{d x^3}
- 6 x (2 + x^2) \frac{d^2}{d x^2}
+ 6 (2 - 3 x^2) \frac{d}{d x}
+  48 x 
 \right] \Phi^{(n)}(x) = K^{(n)}(x) \ , 
\label{eq:eqPhin} 
\end{equation} 
with 
\begin{eqnarray} 
  K^{(n)}(x) &=&
          \bigg\{
            24 x
          + \big( 3 + 24 x^2 \big) \frac{d}{dx}
        \bigg\} \Phi^{(n-3)}(x) 
        \nonumber \\
& &
        + \bigg\{
            92 x
          + \big( 16 + 40 x^2 \big) \frac{d}{dx}
          - \big( x + 26 x^3 \big) \frac{d^2}{dx^2}
        \bigg\} \Phi^{(n-2)}(x)
        \nonumber \\
& &
        + \bigg\{
            116 x
          + \big( 25 - 2 x^2 \big) \frac{d}{dx}
          - \big( 13 x + 32 x^3 \big) \frac{d^2}{dx^2}
          - \big( 3 x^2 - 9 x^4 \big) \frac{d^3}{dx^3}
        \bigg\} \Phi^{(n-1)}(x)
\ , 
\label{eq:Kn} 
\end{eqnarray} 
which shows that the equation at a given order $n$ for $\Phi^{(n)}(x)$ 
involves in the inhomogeneous term the coefficients 
$\Phi^{(k)}(x)$ (and their derivatives) with $k<n$ 
-- hence the ``chained equations" expression used above 
(obviously $\Phi^{(k)}(x) = 0$ when $k < -4$). 
Such a structure calls for an algorithm of solution bottom-up, i.e. 
starting from the lowest value of $n$ (which is $n=-4$) and proceeding 
recursively to the next $n+1$ value up to the required order. 
\par 
The Eq.s(\ref{eq:eqPhin}) have all the same associated homogeneous 
equation, independent of $n$, 
\begin{equation} 
 \left[ 
  x^3 (1-x^2) \frac{d^4}{d x^4}
+ x^2 (1 + 5 x^2) \frac{d^3}{d x^3}
- 6 x (2 + x^2) \frac{d^2}{d x^2}
+ 6 (2 - 3 x^2) \frac{d}{d x}
+  48 x 
 \right] \phi(x) = 0 \ ; 
\label{eq:eqhomophi} 
\end{equation} 
once the solutions of Eq.(\ref{eq:eqhomophi}) are known, all the 
Eq.s(\ref{eq:eqPhin}) can be solved by the method of the 
variation of the constants of Euler. 
\par 
To our (pleasant) surprise, the solutions of Eq.(\ref{eq:eqhomophi}) are 
almost elementary. By trial and error, a first solution is found to be 
\begin{equation} 
 \phi_1(x) = x^2 \ . 
\label{eq:phi1} 
\end{equation} 
We then substitute $\phi(x) = \phi_1(x) \xi(x) $ in Eq.(\ref{eq:eqhomophi}), 
obtaining the following 3rd order equation for the derivative of
$\xi(x) $  
\begin{equation} 
 \left[ 
  x^3 (1 - x^2) \frac{d^3}{d x^3}
+ 3 x^2 (3 - x^2) \frac{d^2}{d x^2}
+ 6 x (1 + 2 x^2) \frac{d}{d x}
- 6 ( 5 + 2 x^2 ) 
 \right] \xi'(x) = 0\ , 
\label{eq:eqxi} 
\end{equation} 
and find that it admits the solution 
\begin{equation} 
 \xi'_2(x) = \frac{1}{x^3}(1-x^2+x^4) \ . 
\label{eq:xi2} 
\end{equation} 
Substituting $\xi'(x) = \xi'_2(x) \chi(x)$ in Eq.(\ref{eq:eqxi}) we 
obtain the following 2nd order equation for $\chi'(x)$
\begin{equation} 
 \left[ 
    x^2 (1 - x^2) (1-x^2+x^4) \frac{d^2}{d x^2}
+ 6 x^5 (2 - x^2) \frac{d}{d x}
- 6 ( 2 - 2 x^4 + x^6 ) 
 \right] \chi'(x) = 0\ , 
\label{eq:eqchi} 
\end{equation} 
which admits as solution 
\begin{equation} 
 \chi'_3(x) = \frac{1}{x^3}(1-x^2)^4\ \frac{5-2x^2+5x^4}{(1-x^2+x^4)^2} \ . 
\label{eq:chi3} 
\end{equation} 
Finally, substituting $ \chi'(x) = \chi'_3(x)\tau(x) $ in 
Eq.(\ref{eq:eqchi}), we obtain the equation 
\begin{eqnarray} 
 \bigg[ 
 x (1 - x^2)^5 (1-x^2+x^4) (5-2x^2+5x^4) \frac{d}{d x}
  \hspace*{1.5cm} & & \nonumber \\
-2 (1 - x^2)^4 (15 - 12x^2 + 11x^4 + 30x^6 - 24x^8 + 20x^{10}) 
 \bigg]& & \hspace*{-0.5cm} \tau'(x) = 0\ , 
\label{eq:eqtau} 
\end{eqnarray} 
which has the solution 
\begin{equation} 
 \tau'_4(x) = \frac{x^6}{(1-x^2)^5}\ \frac{1-x^2+x^4} {5-2x^2+5x^4}\ . 
\label{eq:tau4} 
\end{equation} 

By repeated quadratures in $x$ and multiplications by the previous 
solutions we obtain the explicit analytic expressions of the 4 solutions 
of Eq.(\ref{eq:eqhomophi}); the nasty denominators $(1-x^2+x^4)$ and 
$(5-2x^2+5x^4)$ disappear in the final results, while the repeated 
integrations of the terms with denominators $x, (1+x)$ and $(1-x)$ 
generate, almost by definition, Harmonic PolyLogarithms~\cite{hpl} 
of argument $x$ and weight increasing up to 3. Explicitly, one finds 
\begin{eqnarray}
 \phi_2(x) &=&
       - \frac{1}{2} (1 - x^4)
       - H(0,x) x^2 \ ,  \\ 
      & & \nonumber \\
 \phi_3(x) &=&
      \frac{(5 + 18x^2 + 14x^6 + 5x^8)}{8 \ x^2}
       + \frac{1}{2} (12 + x^2 - 12 x^4 ) H(0;x)
       + 12 \ x^2 H(0,0;x) \ ,  \\ 
      & & \nonumber \\
 \phi_4(x) &=&
          \frac{(1 + x^2) (15 + 182 x^2 + 15 x^4)}{65536 \ x} 
        + \frac{3 (1-x^2)^2 (5 + x^2) (1 + 5x^2)}{131072 \ x^2} 
              \Big[H(-1;x) + H(1;x) \Big] \nonumber \\
      & &   - \frac{9 (1 - x^4)}{8192}  \Big[ H(0,-1;x) + H(0,1;x) \Big] 
            - \frac{9 x^2}{4096}  \Big[ H(0,0,-1;x) + H(0,0,1;x) \Big] \ .
\label{eq:phis} 
\end{eqnarray}
The corresponding Wronskian has the remarkably simple expression 
\begin{eqnarray}
W(x) =  \begin{array}{|c c c c|}
             \phi_1(x)  & \phi_2(x)  & \phi_3(x)  &  \phi_4(x) \\ 
               & & & \\
             \phi_1'(x)  & \phi_2'(x)  & \phi_3'(x)  &  \phi_4'(x) \\
               & & & \\
             \phi_1''(x)  & \phi_2''(x)  & \phi_3''(x)  &  \phi_4''(x) \\
               & & & \\
             \phi_1'''(x)  & \phi_2'''(x)  & \phi_3'''(x)  &  \phi_4'''(x)
        \end{array} = \frac{(1-x^2)^3}{x} \ ,
\label{eq:W} 
\end{eqnarray}
in agreement (of course) with the coefficients of the $4^{th}$ and
$3^{rd}$ $x$-derivative of $\phi(x)$ in Eq.(\ref{eq:eqhomophi}). 

\section{The Solution of the differential equations for the coefficients 
of the expansion in $(d-4)$. } 
\label{sec:solved-4} 
\setcounter{equation}{0} 
With the results established in the previous Section one can 
use Euler's method of the variation of the constants for solving 
Eq.s(\ref{eq:eqPhin}) recursively in $n$, starting from $n=-4$. 
We write Euler's formula as 
\begin{equation} 
 \Phi^{(n)}(x) = \sum_{i=1}^4 \ \phi_i(x) \left[ \Phi_i^{(n)} 
               + \int_0^x \frac{dx'}{W(x')} M_i(x') K^{(n)}(x') \right] \ , 
\label{eq:Euler} 
\end{equation} 
where the $\phi_i(x)$ are the solutions of the homogeneous equation 
given in Eq.s(\ref{eq:phi1},\ref{eq:phis}), the $ \Phi_i^{(n)} $ 
are the as yet undetermined integration constants, the 
Wronskian $W(x)$ can be read from Eq.(\ref{eq:W}), the $M_i(x)$ 
are the minors of the $\phi_i^{'''}(x) $ in the determinant 
Eq.(\ref{eq:W}), and the $K^{(n)}(x)$ 
are the inhomogeneous terms of Eq.(\ref{eq:Kn}). 
The constants $\Phi_i^{(n)}$ are then fixed by comparing the 
expansion in $x$ for $x\to 0$ of Eq.(\ref{eq:Euler}) with the 
expansion in $(d-4)$ for $d\to4$ of Eq.(\ref{eq:Phiexpexpl}). 
Explicitly, we find for instance 
\begin{eqnarray} 
 \Phi^{(-4)}(x) &=& - \frac{1}{64} x^2 \ , 
\label{eq:Phi-4x} \\
 \Phi^{(-3)}(x) &=&  
       - \frac{1}{384}
       + \frac{9}{256} x^2
       - \frac{1}{192} x^4
       - \frac{1}{48} x^2 H(0;x)  \ , 
\label{eq:Phi-3x} \\
 & & \nonumber \\ 
 \Phi^{(-2)}(x) &=&  
         \frac{41}{4608}
       - \frac{143}{3072} x^2
       + \frac{37}{2304} x^4
       - \frac{5}{9216} x^6
       + \bigg( \frac{3}{64} - \frac{1}{96} x^2 \bigg) x^2 H(0;x) 
       - \frac{1}{48} x^2 H(0,0;x) \ ,
\label{eq:Phi-2x} \\
 & & \nonumber \\ 
 \Phi^{(-1)}(x) &=&  
       - \frac{805}{55296}
       + \frac{205}{4096} x^2
       - \frac{743}{27648} x^4
       + \frac{599}{221184} x^6
       - \bigg(
                  \frac{43}{768} 
                - \frac{11}{384} x^2
                + \frac{5}{2304} x^4
         \bigg) x^2 H(0;x)
       \nonumber \\
       & &
       + \bigg( \frac{3}{64} - \frac{1}{96} x^2 \bigg) x^2 H(0,0;x)
       - \frac{1}{48} x^2 H(0,0,0;x) \ ,
\label{eq:Phi-1x} \\
 & & \nonumber \\ 
 \Phi^{(0)}(x) &=&  
         \frac{3173}{663552}
       - \frac{1}{3840} \pi^4 x^2 
       - \frac{2443}{49152} x^2
       - \frac{1}{96} x^4 \zeta(3)
       + \frac{15649}{331776} x^4
       - \frac{34061}{5308416} x^6
       + \frac{1}{96} \zeta(3)
       \nonumber \\
       & &
       - \bigg(
                  \frac{5}{1536} 
                - \frac{1}{48} \zeta(3) x^2 
                + \frac{1}{768}  x^2
                + \frac{121}{2304} x^4
                - \frac{599}{55296} x^6
         \bigg) H(0;x) 
       + \bigg(
                  \frac{1}{96} 
                + \frac{5}{1536} \frac{1}{x^2}
                \nonumber \\
                & &
                - \frac{7}{256}  x^2
                + \frac{1}{96} x^4
                + \frac{5}{1536} x^6
         \bigg) \bigg[ H(1,0;x) - H(-1,0;x) \bigg]
       - \bigg(
                  \frac{43}{768} x^2
                - \frac{11}{384} x^4
                \nonumber \\
                & &
                + \frac{5}{2304} x^6
         \bigg) H(0,0;x)
       + \frac{(1 - x^4)}{32}  \bigg[ H(0,1,0;x) - H(0,-1,0;x) \bigg] 
       + \frac{1}{16}  x^2 \bigg[   H(0,0,1,0;x) 
                                  \nonumber \\
                                  & &
                                  - H(0,0,-1,0;x) \bigg]
       + \bigg(
                  \frac{3}{64} 
                - \frac{1}{96} x^2
         \bigg) x^2 H(0,0,0;x) 
       - \frac{1}{48} x^2 H(0,0,0,0;x) \ .
\label{eq:Phi0x}
\end{eqnarray} 
The full results become quickly too lengthy to be reported explicitly here, 
so we give only the values of the integration constants up to order $3$ 
included. We find $\Phi_4^{(k)} = 0$, for (all) $-4 \le k \le 3$; the 
other constants are 
\begin{eqnarray}
\Phi_1^{(1)} &=&
         \frac{24978775}{603979776} 
       + \frac{32419}{786432} \zeta(3) 
       + \frac{4181}{5242880} \pi^4 
       - \frac{7}{64} \zeta(5) \ ; \\
\Phi_2^{(1)} &=&
       - \frac{19014089}{905969664} 
       + \frac{34481}{393216} \zeta(3)
       - \frac{8507}{7864320} \pi^4 \ ; \\
\Phi_3^{(1)} &=&
         \frac{264575}{452984832} 
       - \frac{871}{196608} \zeta(3)
       - \frac{19}{3932160} \pi^4 \ ; \\
 & & \nonumber \\
\Phi_1^{(2)} &=&
         \frac{1791217627}{7247757312} 
       - \frac{476887}{3145728} \zeta(3) 
       + \frac{37601}{62914560} \pi^4 
       + \frac{87801}{262144} \zeta(5)
       + \frac{1}{32} \zeta^2(3) 
       - \frac{13}{32256} \pi^6 \ ; \\
\Phi_2^{(2)} &=&
         \frac{674733523}{10871635968} 
       - \frac{691111}{4718592} \zeta(3) 
       + \frac{79069}{18874368} \pi^4 
       - \frac{43165}{131072} \zeta(5) \ ; \\
\Phi_3^{(2)} &=&
       - \frac{191322157}{5435817984} 
       + \frac{27041}{2359296} \zeta(3) 
       - \frac{1787}{9437184} \pi^4 
       - \frac{133}{65536} \zeta(5) \ ; \\
 & & \nonumber \\
\Phi_1^{(3)} &=&
       - \frac{40220928899}{21743271936} 
       + \frac{6041939}{9437184} \zeta(3) 
       - \frac{775961}{150994944} \pi^4 
       + \frac{3855}{1048576} \zeta(5) 
       - \frac{12543}{131072} \zeta^2(3) 
       \nonumber \\ 
       & & 
       + \frac{54353}{44040192} \pi^6 
       + \frac{1}{384} \zeta(3) \pi^4  
       - \frac{605}{512} \zeta(7) \ ; \\
\Phi_2^{(3)} &=&
         \frac{336259373}{16307453952} 
       + \frac{1366211}{14155776} \zeta(3) 
       - \frac{7536877}{1132462080} \pi^4 
       + \frac{1939871}{1572864} \zeta(5) 
       + \frac{15281}{196608} \zeta^2(3) 
       \nonumber \\ 
       & & 
       - \frac{69631}{66060288} \pi^6 \ ; \\
\Phi_3^{(3)} &=&
         \frac{3741658007}{16307453952} 
       - \frac{381061}{7077888} \zeta(3) 
       + \frac{275627}{566231040} \pi^4 
       - \frac{41641}{786432} \zeta(5)
       + \frac{19}{32768} \zeta^2(3) 
       \nonumber \\ 
       & & 
       - \frac{247}{33030144} \pi^6 \ ; 
\label{eq:Phini} 
\end{eqnarray} 
\begin{figure}[h] 
\begin{center} 
\includegraphics*[2cm,1cm][14cm,5cm]{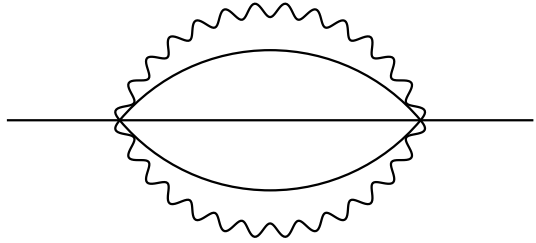} 
\caption{The on-shell 4-loop sunrise graph.} 
\end{center}
\label{fig:2} 
\end{figure} 

As a byproduct, using the results of ~\cite{VermTab}, we obtain the 
values at $x=1$, i.e. the on mass-shell values at $-p^2=m^2=M^2$ 
of Eq.(\ref{eq:defFi}), depicted in Fig.2, 
\begin{eqnarray} 
\Phi(d,x=1) &=&
           - \frac{1}{64 \ (d-4)^4 }
           + \frac{7}{256 \ (d-4)^3 }
           - \frac{17}{768 \ (d-4)^2 }
           + \frac{835}{73728 \ (d-4) }
           - \frac{7379}{1769472}
           - \frac{1}{1920} \pi^4
           \nonumber \\
     & &
           + (d-4) \bigg\{
                           - \frac{6766055}{42467328}
                           + \frac{53}{1152} \pi^2
                           + \frac{7}{7680} \pi^4
                           - \frac{7}{256} \pi^2 \zeta(3)
                           + \frac{29}{256} \zeta(5)
                   \bigg\}
           \nonumber \\
     & &
           + (d-4)^2 \bigg\{
                               \frac{1449210865}{1019215872}
                             - \frac{1913}{4608} \pi^2
                             + \frac{53}{128} \pi^2 \ln2
                             - \frac{1855}{2304} \zeta(3)
                             - \frac{17}{23040} \pi^4
                             + \frac{49}{1024} \pi^2 \zeta(3)
                             \nonumber \\
                             & & \qquad \qquad \quad
                             - \frac{203}{1024} \zeta(5)
                             - \frac{63}{256} \pi^2 \zeta(3) \ln2 
                             - \frac{9}{32} a_4 \pi^2 
                             - \frac{3}{256} \pi^2 \ln^4 2
                             + \frac{3}{256} \pi^4 \ln^2 2
                             \nonumber \\
                             & & \qquad \qquad \quad
                             + \frac{245}{1024} \zeta^2(3)
                             + \frac{661}{322560} \pi^6
                   \bigg\}
           \nonumber \\
     & &
           + (d-4)^3 \bigg\{
          - \frac{ 182188906799}{24461180928}
          + \frac{ 734695}{331776} \pi^2
          - \frac{ 1913}{512} \pi^2 \ln2
          + \frac{ 66955}{9216} \zeta(3)
          + \frac{ 583}{384} \pi^2 \ln^2 2
          \nonumber \\
          & & \qquad \qquad \quad
          - \frac{ 8707}{147456} \pi^4
          + \frac{ 265}{768} \ln^4 2
          + \frac{ 265}{32} a_4
          - \frac{ 119}{3072} \pi^2 \zeta(3)
          + \frac{ 493}{3072} \zeta(5)
          + \frac{ 441}{1024} \pi^2 \zeta(3) \ln2 
          \nonumber \\
          & & \qquad \qquad \quad
          + \frac{ 21}{1024} \pi^2 \ln^4 2
          - \frac{ 21}{1024} \pi^4 \ln^2 2
          - \frac{ 661}{184320} \pi^6
          + \frac{ 63}{128} a_4 \pi^2 
          - \frac{ 1715}{4096} \zeta^2(3)
          \nonumber \\
          & & \qquad \qquad \quad
          - \frac{ 627}{512} \pi^2 \zeta(3) \ln^2 2 
          - \frac{ 123}{1280} \pi^2 \ln^5 2
          + \frac{ 9917}{4096} \pi^2 \zeta(5)
          + \frac{ 107}{1536} \pi^4 \ln^3 2
          \nonumber \\
          & & \qquad \qquad \quad
          - \frac{ 109}{12288} \pi^4 \zeta(3)
          + \frac{ 2205}{1024}  \zeta^2(3) \ln2
          + \frac{ 45}{128}  \zeta(3) \ln^4 2
          - \frac{ 15}{2048} \pi^6 \ln2
          \nonumber \\
          & & \qquad \qquad \quad
          - \frac{ 81}{32} a_4 \pi^2  \ln2 
          + \frac{ 45}{16} a_4 \zeta(3) 
          - \frac{ 81}{32} a_5 \pi^2 
          - \frac{ 45}{16} a_5 \ln^2 2 
          + \frac{ 1395}{1024} \zeta(5) \ln^2 2 
          \nonumber \\
          & & \qquad \qquad \quad
          - \frac{ 45}{4} a_6 \ln2 
          + \frac{ 45}{16} b_6 \ln2 
          - \frac{ 135}{8} a_7
          + \frac{ 45}{8} b_7
          - \frac{ 45}{32} d_7
          + \frac{ 15}{1792} \ln^7 2
          + \frac{ 29335}{2048} \zeta(7)
                     \bigg\} \nonumber \\
& & + {\mathcal O}\Big( (d-4)^4 \Big) \ .
\label{eq:Phinx=1} 
\end{eqnarray} 
At variance with~\cite{VermTab}, we have expressed the results in terms 
of the constants listed in Table 1; the first column is the name of the 
constant, the second column its value as Harmonic PolyLogarithm of 
suitable argument, the third as Nielsen PolyLogarithm (when available), 
the last the numerical value. 

\begin{table}[!h]
\begin{center}
\begin{tabular}{|c|c|c|l|}
\hline
 constant & HPL & NPl & numerical value \\
\hline
\hline
\hline
 $ \zeta(3) $ & $ H(0,0,1;1) $ & $ S_{2,1}(1) $
              & 1.2020569031595942854 \\
 $ a_4 $      & $ H(0,0,0,1;1/2) $ & $ S_{3,1}(1/2) $
              & 0.51747906167389938633 \\
 $ \zeta(5) $ & $ H(0,0,0,0,1;1) $ & $ S_{4,1}(1) $
              & 1.0369277551433699263 \\
 $ a_5 $      & $ H(0,0,0,0,1;1/2) $   & $ S_{4,1}(1/2) $
              & 0.50840057924226870746 \\
 $ a_6 $      & $ H(0,0,0,0,0,1;1/2) $ & $ S_{5,1}(1/2) $
              & 0.50409539780398855069 \\
 $ b_6 $      & $ H(0,0,0,0,1,1;1/2) $ & $ S_{4,2}(1/2) $
              & 0.0087230030575968884272  \\
 $ \zeta(7) $ & $ H(0,0,0,0,0,0,1;1) $ & $ S_{6,1}(1) $
              & 1.0083492773819228268 \\
 $ a_7 $      & $ H(0,0,0,0,0,0,1;1/2) $   & $ S_{6,1}(1/2) $
              & 0.50201456332470849457 \\
 $ b_7 $      & $ H(0,0,0,0,0,1,1;1/2) $ & $ S_{5,2}(1/2) $
              & 0.0041965726953603256975  \\
 $ d_7 $      & $ H(0,0,0,0,1,-1,-1;1) $ & $ -- $
              & 0.0022500546439578516764 \\
\hline
\end{tabular}
\end{center}
\caption{Constants up to weight 7 appearing in Eq.(\ref{eq:Phinx=1}) }
\label{tab:const}
\end{table}

Once the explicit analytic expressions of the coefficients of the 
Laurent-expansion of $\Phi(d,x)$ in $(d-4)$ are known, one can use 
Eq.s(\ref{Phi2fromPhi}-\ref{Phi4fromPhi}) for obtaining the coefficients 
of $\Phi_2(d,x), \Psi_3(d,x)$ and $\Phi_4(d,x)$. 
The coefficients of the expansion of $\Psi_5(d,x)$ are then 
recovered by integrating in $x$ Eq.(\ref{Psi5eq}); the quadrature 
is trivial to carry out in terms of HPLs, and the 
integration constants are fixed by the condition $\Psi_5(d,0)=0$, which 
follows at once from 
Eq.s(\ref{defPsi5}),(\ref{eq:FtoPhi}),(\ref{eq:defFi}). 
\section{The Independent numerical calculation.} 
\label{sec:numcal} 
\setcounter{equation}{0} 
In this Section we will calculate $\Phi(d,x=1)$ with high numerical 
precision by suitably using the difference equation method described 
in~\cite{lap2,lap} and references therein. 
At variance with ~\cite{lap}, we will apply the method directly to the 
Master Integral with massless lines. This will imply some additional 
work for establishing the initial conditions of the difference equations. 

We define 
\begin{equation} 
\label{eq:defI5} 
I_{5}(d,n)= \pi^{-2d} \int \frac{d^dk_1\  d^dk_2 \ d^dk_3 \ d^dk_4 } 
      { (k_1^2+1)^n ((k_2-k_1)^2+1) ((k_3-k_2)^2+1) (k_4-k_3)^2 (p-k_4)^2 }
      \ , \quad p^2=-1\ ,  
\end{equation} 
so that $I_5(d,1)$ is equal to $\Phi(d,x)$ of Eq.(\ref{eq:defPhi}) 
at $x=1$ up to a known multiplicative factor  
\begin{equation}\label{eq:normal}
I_{5}(d,1) = {\left[4\Gamma(1+\ep)\right]^4} \Phi(d=4-2\ep,x=1)\ .
\end{equation}
By combining identities obtained by integration by parts 
one finds that $I_{5}(d,n)$ satisfies the third-order difference equation 
\begin{eqnarray} 
 32(n-1)(n-2)(n-3)(n-3d+5)I_{5}(d,n) & & \nonumber\\ 
 -4(n-2)(n-3)\biggl[15n^2+(39-50d)n +27d^2 +5d -54 
                \biggr] I_{5}(d,n-1) & & \nonumber\\ 
 +2(n-3) \biggl[12n^3-(38d+24)n^2+(23d^2+133d-84)n+9d^3-141d^2 
                      +134d-24 \biggr] I_{5}(d,n-2) & & \nonumber\\ 
 +(n-d-1)(n-2d+1)(2n-3d)(2n-5d+4)I_{5}(d,n-3) = 0 && \;.
\label{eq:dif1} 
\end{eqnarray}
We will solve this difference equation by using the Laplace {\it ansatz} 
\begin{equation}\label{eq:ansatz5} 
I_{5}(d,n)=\int_0^1 t^{n-1} v_{5}(d,t) \; dt\ , 
\end{equation} 
giving for $v_{5}(d,t)$ the fourth-order differential equation
\begin{eqnarray} 
 && 4t^4(8t+1)(t-1)^2 \frac{d^4}{dt^4}v_5(d,t)     \nonumber\\ 
 && +4t^3(t-1) \biggl[ 24(d+1)t^2 +(18 - 26d)t - 7d + 12 \biggr] 
                          \frac{d^3}{dt^3}v_5(d,t) \nonumber\\
 && t^2\biggl[ (576 (d-1)t^3 + ( - 108d^2 - 420d +648   )t^2 
      +(46d^2  + 38d  - 144 )t  + 71d^2  - 284d   + 288 \biggr] 
                          \frac{d^2}{dt^2}v_5(d,t) \nonumber\\
 && +t\biggl[(576d-960)t^3 +(- 216d^2+360d)t^2  \biggr. \nonumber\\ 
 && \phantom{t[} \biggl. +(- 18d^3 + 190d^2 - 496d +384)t 
                 + 77d^3  - 533d^2   + 1236d - 960) \biggr] 
                                   \frac{d}{dt}v_5(d,t) \nonumber\\
 && + (d-3)(2d-5)(3d-8)(5d-12) v_5(d,t) =0\;. 
\label{eq:equd1} 
\end{eqnarray} 
We will look for the solution of Eq.(\ref{eq:equd1}) in the form of 
a power series expansions; inserted in Eq.(\ref{eq:ansatz5}) 
and integrated term by term it will provide very accurate values 
of $I_5(d,n)$. As the convergence is faster for larger $n$, we will 
consider ``large enough" values of the index $n$ (see below); the repeated 
use ``top-down" of Eq.(\ref{eq:dif1}) ({\it i.e.} using it for 
expressing $I_5(d,n-3)$ in terms of the $I_5(k)$ with $k=n,n-1,n-2$) 
will give the values corresponding to smaller indices, till $I_5(d,1)$ 
is eventually obtained. 
To go on with this program, initial conditions for $v_5(d,t)$ are needed. 

From the definition Eq.(\ref{eq:defI5}), and introducing spherical 
coordinates in $d$-dimension for the loop momentum $k_1$, 
$ d^dk_1 = k_1^{d-1}dk_1 d\Omega(d,\hat k_1) $ one has 
\begin{eqnarray}
I_{5}(d,n)&=&\dfrac{1}{\Gamma\left(\frac{d}{2}\right)} 
    \int_0^\oo \dfrac{(k_1^2)^{d/2-1}\;dk_1^2}{(k_1^2+1)^n} 
      f_{5}(d,k_1^2) \;, \\
\label{eq:deff2}
f_{5}(d,k_1^2)&=& \int \dfrac{d\Omega(d,\hat k_1)}{\Omega(d)} 
                         I_{4}(d,1,(p-k_1)^2) \;,
\end{eqnarray} 
where $\Omega(d)$ is the $d$-dimensional solid angle,
and $I_{4}(d,1,(p-k_1)^2) $ is the 3-loop (off mass-shell) sunrise integral 
\begin{equation}\label{eq:defI4} 
I_{4}(d,n,(p-k_1)^2)= \pi^{-3d/2} \int \frac{d^dk_2\  d^dk_3 \ d^dk_4} 
      { ((k_2-k_1)^2+1)^n ((k_3-k_2)^2+1) (k_4-k_3)^2 (p-k_4)^2  }
      \ . 
\end{equation} 
By the change of variable $1/(k_1^2+1)=t,\ $ $\ k_1^2=(1-t)/t,\ $ one finds
\begin{equation}
  I_{5}(d,n)=\dfrac{1}{\Gamma\left(\frac{d}{2}\right)} 
             \int_0^1 t^{n-1} (1-t)^{\frac{d}{2}-1} 
             t^{-\frac{d}{2}} f_{5}\left(d,\frac{1-t}{t}\right) \; dt\;,
\end{equation}
from which one gets the relation between $v_{5}(d,t) $ and 
$f_{5}(d,(1-t)/t)$
\begin{equation}\label{eq:vvff}
v_{5}(d,t)=\dfrac{1}{\Gamma\left(\frac{d}{2}\right)} 
             (1-t)^{\frac{d}{2}-1} 
             t^{-\frac{d}{2}} f_{5}\left(d,\frac{1-t}{t}\right) \; dt\;. 
\end{equation} 
From that relation we see that we can derive boundary conditions for 
$v_5(d,t)$ in the $t\to1$ limit from the expansion of $f_5(d,k_1^2)$ 
in the $k_1\to 0 $ limit, which is easy to obtain. 
Indeed, only the first denominator of \eqref{eq:defI4} depends on $k_1$; 
expanding for small $k_1$ and performing the angular integration one gets
\begin{equation}
\begin{split}
\int \dfrac{d\Omega(d,\hat k_1)}{\Omega(d)} \dfrac{1}{(k_2-k_1)^2+1}
=& \int \dfrac{d\Omega(d,\hat k_1)}{\Omega(d)} \left(
\dfrac{1}{k_2^2+1} 
-\dfrac{k_1^2-2k_1\cdot k_2}{(k_2^2+1)^2} 
+\dfrac{(k_1^2-2k_1\cdot k_2)^2}{(k_2^2+1)^3} + {\ldots} \; \right)\\
=&\dfrac{1}{k_2^2+1} 
+k_1^2\left(-\dfrac{1}{(k_2^2+1)^2} +\dfrac{4}{d} 
                        \dfrac{k_2^2}{(k_2^2+1)^3}\right) + {\ldots} \;.
\end{split}
\end{equation}
The above result gives the expansion of $f_{5}(d,k_1^2)$ at $k_1^2=0$:
\begin{equation}\label{eq:svilf5}
\begin{split}
 f_{5}(d,k_1^2)\ =& \ f_5^{(0)}(d) + f_5^{(1)}(d)\ k_1^2 + O(k_1^4)\;, \\
   f_5^{(0)}(d)\ =& \ I_{4}(d,1,p^2) \\ 
   f_5^{(1)}(d)\ =& \ -I_{4}(d,2,p^2)
  +\dfrac{4}{d}\left[I_{4}(d,2,p^2) - I_{4}(d,3,p^2) \right] \ . \\ 
\end{split} 
\end{equation}
Note that $f_5(d,k_1^2)$ is regular in the origin. 

By inspecting the differential equation \itref{eq:equd1} 
one finds that the behaviour at $t=1$ of the 4 independent solution
is $\sim(1-t)^{\alpha_i}$, with $\alpha_1=d/2-1$, $\alpha_2=d/2$, 
$\alpha_3=0$, and $\alpha_4=1$; for comparison with \eqref{eq:vvff} 
the behaviours $\alpha_3=0$, and $\alpha_4=1$ are ruled out and the 
expansion reads 
\begin{equation}\label{eq:cond0} 
v_{5}(d,t)= (1-t)^{\frac{d}{2}-1}\left( v_5^{(0)}(d) 
                   + v_5^{(1)}(d)\ (1-t) +O(1-t)^2 \right) \;; 
\end{equation}
by comparison with \eqref{eq:svilf5} ($t=1$ corresponds to $k_1^2=0$), 
one obtains 
\begin{equation}\label{eq:cond1} 
\begin {split} 
v_5^{(0)}(d) = &\frac{1}{\Gamma\left(\frac{d}{2}\right)} f_5^{(0)}(d) \;, \\ 
v_5^{(1)}(d) = &\frac{1}{\Gamma\left(\frac{d}{2}\right)} 
         \left[ \frac{d}{2}f_5^{(0)}(d) + f_5^{(1)}(d) \right] \;.
\end{split}
\end{equation} 
The values $I_{4}(d,n)$ of $I_{4}(d,n,p^2)$ at $p^2=-1$ are therefore 
required 
\begin{equation}\label{eq:defI4b} 
I_{4}(d,n)\equiv
I_{4}(d,n,p^2)=
\pi^{-3d/2} \int \frac{d^dk_2\  d^dk_3 \ d^dk_4} 
      { (k_2^2+1)^n ((k_3-k_2)^2+1) (k_4-k_3)^2 (p-k_4)^2  }, \quad p^2=-1
      \ . 
\end{equation} 
The problem of evaluating the $I_{4}(d,n)$ is similar to the 
original problem of evaluating the $I_{5}(d,n)$, but in fact it 
is much simpler, as the $I_{4}(d,n)$ involve one less loop and one 
less propagator. 
As above, by using integration-by-parts identities one finds that 
$I_{4}(d,n)$ satisfies the third-order difference equation 
\begin{eqnarray} 
 && 6(n-1)(n-2)(n-3)I_{4}(d,n) \nonumber\\ 
 && -(n-2)(n-3)(10n-7d-10)I_{4}(d,n-1) \nonumber\\ 
 && +(n-3)(2n^2+(2d-18)n-7d^2+29d-8)I_{4}(d,n-2) \nonumber\\ 
 && +(n-d-1)(n-2d+1)(2n-3d)I_{4}(d,n-3)=0\;.
\label{eq:dif2}
\end{eqnarray} 
We solve the difference equation by using again the Laplace {\it ansatz} 
\begin{equation}\label{eq:ansatz4} 
I_{4}(d,n)=\int_0^1 t^{n-1} v_{4}(d,t) \; dt\;,
\end{equation} 
where $v_{4}(d,t)$ satisfies the differential equation
\begin{eqnarray} 
 && 2t^3(3t+1)(t-1)^2  \frac{d^3}{dt^3}v_{4}(d,t) \nonumber\\ 
 && +t^2(t-1)(36t^2+(6-7d)t - 9d + 18 ) \frac{d^2}{dt^2}v_{4}(d,t) \nonumber\\
 && +t(36t^3-14dt^2 +(-7d^2+33d-36)t + 13d^2- 61d + 72) 
                                        \frac{d}{dt}v_{4}(d,t) \nonumber\\
 && +(d-3)(2d-5)(3d-8) v_{4}(d,t) =0\;.
\label{eq:equd2} 
\end{eqnarray} 

Following the procedure used above, we write
\begin{eqnarray}
I_{4}(d,n)&=&\dfrac{1}{\Gamma\left(\frac{d}{2}\right)} 
  \int_0^\oo\dfrac{(k_2^2)^{d/2-1}\; dk_2^2}{(k_2^2+1)^n} f_{4}(k_2^2) \;, \\
\label{eq:deff3} 
f_{4}(k_2^2)&=&\int \dfrac{d\Omega(d,\hat k_2)}{\Omega(d)} I_{3}(d,(p-k_2)^2)
\;, \\
\label{eq:defI3}
I_{3}(d,(p-k_2)^2)&=& \pi^{-d} \int \frac{d^dk_3 \ d^dk_4} 
      { ((k_3-k_2)^2+1) (k_4-k_3)^2 (p-k_4)^2  }
      \ , \\
\label{eq:vvff2}
v_{4}(d,t)&=&\dfrac{1}{\Gamma\left(\frac{d}{2}\right)} 
           (1-t)^{\frac{d}{2}-1} t^{-\frac{d}{2}} 
            f_{4}\left(d,\frac{1-t}{t}\right) \; . 
\end{eqnarray} 
At variance with the previous case, the function $f_{4}(d,k_2^2)$ is 
\emph{not} regular for $k_2\to 0$, as at $k_2=0$ the value 
of the external momentum squared $(p-k_2)^2$ becomes 
the threshold of the 2-loop sunrise graph associated to $I_{3}(d,p^2)$. 
But it is not difficult to evaluate 
analytically $I_{3}(d,q^2)$ for generic off-shell $q^2$ by using 
Feynman parameters:
\begin{equation*}
I_{3}(d,q^2)
=\dfrac{2\Gamma(5-d)\Gamma\left(3-\frac{d}{2}\right) 
        \Gamma^2\left(\frac{d}{2}-1\right) } 
       {(d-4)^2(3-d)\Gamma\left(\frac{d}{2}\right) } 
    \,{}_2F_1\left(3-d,2-\frac{d}{2};\frac{d}{2};-q^2\right)\ ,
\end{equation*}
where ${}_2F_1$ is the Gauss hypergeometric function.
The expansion of $I_{3}(d,q^2)$ in $q^2=-1$ consists of the sum of two series, 
\begin{equation}\label{eq:I3svil} 
 I_{3}(d,q^2)=  a_0(d) \biggl[1+O(q^2+1)\biggr]
  \ \ \ +\ b_0(d)\ (q^2+1)^{2d-5} \biggl[1+O(q^2+1)\biggr] \ ,
\end{equation}
\begin{equation*}
\begin{split}
a_0(d)\ =\ I_{3}(d,-1)\ =&\ \frac{2\Gamma(5-d)\Gamma\left(3-\frac{d}{2}\right) 
                                \Gamma^2\left(\frac{d}{2}-1\right) 
                    \Gamma(2d-5)}{(4-d)^2(3-d)\Gamma\left(\frac{3}{2}d-3\right) 
                                  \Gamma(d-2)}\;,\\
b_0(d)\ =&\ \Gamma^2\left(\frac{d}{2}-1\right)\Gamma(5-2d)\ .
\end{split}
\end{equation*}
Inserting \eqref{eq:I3svil} into \eqref{eq:deff3}, setting
$q=p-k_2$ and  performing the angular integration over $\hat k_2$
by means of the formula (see Eq.(88) of Ref.\cite{lap2})
valid for $k_2\to 0 $
\begin{equation}\label{eq:hypsvil0}
\dfrac{1}{\Omega(d)}\int \dfrac{d\Omega(d,\hat k_2)}{((p-k_2)^2+1)^N} \approx 
(k_2^2)^{-\frac{N}{2}} 
\dfrac{\Gamma\left(\frac{d}{2}\right) 
       \Gamma\left(\frac{N}{2}\right)} 
     {2\Gamma(N)\Gamma\left(\frac{1}{2}(d-N)\right)}\ ,\quad k_2 \to 0 \ ; 
\end{equation}
with $N=-(2d-5)$, as in the term $(q^2+1)^{2d-5}$ of \eqref{eq:I3svil}, 
one obtains 
\begin{equation}\label{eq:f4svil}
f_{4}(d,k_2^2)= a_0(d)\ \biggl[1+O(k_2^2)\biggr] \ + \ 
  \dfrac{\Gamma\left(\frac{d}{2}\right) 
         \Gamma\left(\frac{5}{2}-d\right)} 
       {2\Gamma(5-2d)\Gamma\left(\frac{1}{2}(3d-5)\right)} \ 
       b_0(d)\ (k_2^2)^{\frac{1}{2}(2d-5)}
       \ \biggl[1+O(k_2^2)\biggr] \;.
\end{equation}
Using the variable $1/(k_2^2+1)=t$ in \eqref{eq:f4svil} and inserting it
in \eqref{eq:vvff2} one gets  
the initial condition for $v_{4}(d,t)$ at the singular point $t=1$
\begin{equation}\label{eq:cond2} 
v_{4}(d,t)= \dfrac{a_0(d)}{\Gamma\left(\frac{d}{2}\right)} 
            (1-t)^{\frac{d}{2}-1} \biggl[1+O(1-t)\biggr] 
 + \dfrac{\Gamma\left(\frac{5}{2}-d\right)} 
   {2\Gamma(5-2d)\Gamma\left(\frac{1}{2}(3d-5)\right)} b_0(d) 
   (1-t)^{\frac{1}{2}(3d-7)} \biggl[1+O(1-t)\biggr] 
\;.
\end{equation}
By inspecting the equation \itref{eq:equd2} 
one gets that the behaviour at $t=0$ of $v_4(d,t)$  is 
\begin{equation}\label{eq:v4at0} 
v_{4}(d,t\to 0)\approx c_4^{(1)}(d)\ t^{-d+3} 
                      +c_4^{(2)}(d)\ t^{-2d+5} 
                      +c_4^{(3)}(d)\ t^{\frac{1}{2}(-3d+8)} \;,
\end{equation}
so that for $d\to 4$ the integral \itref{eq:ansatz4} is convergent 
for $n\ge4$. 

All the quantities depending on $d$ are then systematically expanded 
in $d-4$, the series are truncated at some fixed number of 
terms, and the calculations with the truncated series are performed 
by using the program {\tt SYS}\cite{lap2}; 
as the first 12 terms of the series are lost in the intermediate steps 
of the calculations, in order to obtain the final results, from 
$1/\epsilon^4$ up to $O(\ep^6)$, $23$ initial terms are needed.
We solve finally the differential equation \itref{eq:equd2} 
with the initial condition \itref{eq:cond2} 
by a first expansions in series at $t=1$; due to the presence 
in \eqref{eq:equd2} of a singular point at $t=-1/3$, 
to have fast convergence 
till $t=0$ we switch to subsequent series expansions 
at the intermediate points $1/2$, $1/4$, $1/8$ and $0$; 
then we calculate the integral \itref{eq:ansatz4} 
for $n=4,5,6,7,8$ by integrating the series term by term 
(about 300 terms are needed to reach a precision of 77 digits).
By applying repeatedly ``top-down" the recurrence relation \itref{eq:dif2} 
to $I_{4}(d,8)$, $I_{4}(d,7)$, $I_{4}(d,6)$,
we obtain $I_{4}(d,5)$ and $I_{4}(d,4)$ (which are cross-checked with the
values obtained by direct integration), then $I_{4}(d,3)$, $I_{4}(d,2)$ 
and $I_{4}(d,1)$ 
\begin{multline} \label{eq:resuc}
I_{4}(d,1)=  \Gamma(1+\ep)^3   \biggl[
 0.3333333333333333333333333333333\ep^{-3} 
+\ep^{-2} \\
+1.3611111111111111111111111111111\ep^{-1} 
+0.4254662447518595926183272929396 \\
-31.041066530239171582155933528974\ep 
-110.36756287612408836668594378878\ep^2 \\
-611.01919192086881174365300012691\ep^3 
-1734.0854215005636534710373849833\ep^4 \\
-7316.7112322583252172396619274906\ep^5
 +O(\ep^6) \biggr]\ . 
\end{multline}
Those values of $I_4(d,n)$ are used to determine the initial condition
for $v_5(d,t)$, Eq.s(\ref{eq:cond0},\ref{eq:cond1},\ref{eq:svilf5}).
We then solve the differential equation \itref{eq:equd1}
by expansions in series centered in the points
$t=1$, $1/2$, $1/4$, $1/8$, $1/16$ and $0$
(as above, this subdivision is due to the presence of a singular point at $t=-1/8$).
By inspecting the equation \itref{eq:equd1}
one gets that the behaviour at $t=0$ of
$v_4(d,t)$  is 
\begin{equation}\label{eq:v5at0} 
v_{5}(d,t\to 0)\approx c_5^{(1)}(d) t^{-d+3} 
                      +c_5^{(2)}(d) t^{-2d+5} 
                      +c_5^{(3)}(d) t^{(-3d+8)/2}
                      +c_5^{(4)}(d) t^{(-5d+12)/2}\;,
\end{equation}
so that for $d\to 4$ the integral \itref{eq:ansatz5} is surely convergent
for $n\ge5$; 
then we calculate the integral \itref{eq:ansatz5}
for $n=5,6,7,8,9$ by integrating the series term by term. 
By using repeatedly ``top-down" the recurrence relation \itref{eq:dif2}
starting from $n=9$, we obtain $I_{5}(d,6)$, $I_{5}(d,5)$ (used for 
cross-check), $I_5(d,4),{\ldots}$, $I_{5}(d,1)$. 
The result, up to the coefficient of order 5 in $(d-4)$ included, is 
\begin{multline} \label{eq:resu1}
I_{5}(d,1)=  \Gamma(1+\ep)^4   \biggl[
     -0.25 \ep^{-4}  -0.875 \ep^{-3}  
 -1.416666666666666666666666666667 \ep^{-2} \\ 
 -1.449652777777777777777777777778 \ep^{-1} 
 -14.055442461941065705599451765901 \\
    -90.416062304531327135375791542063 \ep 
    -1170.3684076603804614545918785105 \ep^2 \\
    -5299.6462727245240600241060624284 \ep^3 
    -37132.219579420859394978093377604 \ep^4 \\
    -144381.92488313453838475109116166 \ep^5
 +O(\ep^6) \biggr]
\end{multline}
Taking into account the normalization \itref{eq:normal}
one finds that the first $8$ terms of \eqref{eq:resu1} 
agree with Eq.(\ref{eq:Phinx=1}).

We want only to mention that we have also independently checked the 
numerical result \itref{eq:resu1} by calculating the master integral 
with all masses equal to one by difference equations, and then by using 
the value so obtained as initial condition for the integration of a 
differential equation in the photon mass $\lambda$ from $\lambda=1$ to 
$\lambda=0$.
\section{The $x\to0$ values. } 
\label{sec:xto0val} 
\setcounter{equation}{0} 
We evaluate in this section the $x\to0$ values of $\Phi(d,x)$, 
Eq.(\ref{eq:defPhi}). \par 
By combining the familiar formulae 
\begin{equation} 
  \frac{1}{[(k-l)^2+a)]^\alpha]}\ \frac{1}{(l^2+b)^\beta} = 
  \frac{\Gamma(\alpha+\beta)}{\Gamma(\alpha)\Gamma(\beta)} 
  \int_0^1dx\ \frac{x^{\alpha-1}(1-x)^{\beta-1}} 
      {[(l-xk)^2+x(1-x)k^2+ax+b(1-x)]^{\alpha+\beta}} \ , 
\label{eq:1fp} 
\end{equation} 
and 
\begin{equation} 
  C^{-1}(d) \int \frac{d^dl}{(2\pi)^{d-2}}\ \frac{1}{[(l-q)^2+c]^n} 
   = \frac{1}{4}\ \frac{ \Gamma\left(n-\frac{d}{2}\right) } 
                       { \Gamma\left(3-\frac{d}{2}\right)\Gamma(n) } 
                \ \frac{1}{c^{n-\frac{d}{2}}} \ , 
\label{eq:tadpole} 
\end{equation} 
one gets 
\begin{equation}
 C^{-1}(d) \int \frac{d^dl}{(2\pi)^{d-2}}\ 
  \frac{1}{[(k-l)^2+a]^\alpha}\ \frac{1}{(l^2+b)^\beta} = \frac{1}{4} 
  \frac{ \Gamma\left(\alpha+\beta-\frac{d}{2}\right) } 
      { \Gamma(\alpha)\Gamma(\beta)\Gamma\left(3-\frac{d}{2}\right)} 
  \int_0^1 dx\ \frac{ x^{\frac{d}{2}-\beta-1} (1-x)^{\frac{d}{2}-\alpha-1} } 
                    { [k^2 + d(a,b,x)]^{\alpha+\beta-\frac{d}{2}} } \ , 
\label{eq:1loop} 
\end{equation} 
where 
\[ d(a,b,x) = \frac{ax+b(1-x)}{x(1-x)} \ . \] 
We rewrite Eq.(\ref{eq:defPhi}) as 
\begin{equation} 
  \Phi(d,x) = \frac{C^{-4}(d)}{(2\pi)^{4(d-2)}} 
     \int \frac{d^dq}{[(p-q)^2+x^2]} 
     \int \frac{d^dk_2}{(q-k_2)^2} 
     \int \frac{d^dk_1}{(k_2-k_1)^2} 
     \int \frac{d^dl}{[(k_1-l)^2+1](l^2+1)} \ , 
\label{eq:newdefPhi} 
\end{equation} 
%
and then use repeatedly Eq.(\ref{eq:1loop}), using in the order the parameters 
$ y,y_1,y_2,z $ for integrating the loops $l,k_1,k_2,q$, obtaining 
\begin{eqnarray} 
   \Phi(d,x) &=& \frac{ \Gamma(1-2(d-4))} { 1024\ (d-3)(d-4)(2d-5)(2d-7)
     \Gamma^4\left(1-\frac{1}{2}(d-4)\right) } \nonumber\\ 
 &\times& \int_0^1dy\ y^{\frac{d}{2}-2}(1-y)^{\frac{d}{2}-2} 
 \int_0^1dy_1\ y_1^{\frac{d}{2}-2}(1-y_1)^{d-3} 
 \int_0^1dy_2\ y_2^{\frac{d}{2}-2}(1-y_2)^{\frac{3}{2}d-4} \nonumber\\ 
 &\times& \Psi(d,x,y,y_1,y_2) \ , 
\label{eq:newPhi1} 
\end{eqnarray} 
where 
\begin{equation} 
   \Psi(d,x,y,y_1,y_2) = \int_0^1 dz\ \frac{ z^{3-\frac{3}{2}d} } 
                            { [(1-z)^2x^2+zD(y,y_1,y_2)]^{5-2d} } \ , 
\label{eq:defPsi} 
\end{equation} 
and 
\begin{equation} 
D(y,y_1,y_2) = \frac{1}{y(1-y)(1-y_1)(1-y_2)} \ . 
\label{eq:defD} 
\end{equation} 
The above formulae are valid for any $x$; form now on we take $0<x\ll 1$. 
For definiteness, we take also $d$ to be ``just bigger" than 2 (i.e. 
$d=2+\eta$, with $0 < \eta \ll 1$). 
The $z$-integral will be carried out first. To that aim, introduce 
an infinitesimal parameter $Z$, such that $0<x^2\ll Z\ll 1$ (a possible choice 
might be $Z=-x^2\ \ln{x}$, but the exact value of $Z$ will be irrelevant), 
and split the integration interval as 
\[ \int_0^1 dz = \int_0^Z dz + \int_Z^1 dz \ ; \] 
correspondingly, we write the $z$-integral as 
\begin{eqnarray} 
  \Psi(d,x,y,y_1,y_2) &=& \Psi_1(d,x,y,y_1,y_2) + \Psi_2(d,x,y,y_1,y_2) 
                                                 \ , \label{eq:defPsii} \\ 
  \Psi_1(d,x,y,y_1,y_2) &=& 
        \int_0^Z dz\ \frac{ z^{3-\frac{3}{2}d} } 
           { [(1-z)^2x^2+zD(y,y_1,y_2)]^{5-2d} } \ , \nonumber\\ 
  \Psi_2(d,x,y,y_1,y_2) &=& 
        \int_Z^1 dz\ \frac{ z^{3-\frac{3}{2}d} } 
           { [(1-z)^2x^2+zD(y,y_1,y_2)]^{5-2d} } \ . \nonumber 
\end{eqnarray} 
In the second term we can neglect $x^2$ in the 
denominator obtaining simply 
\begin{eqnarray} 
 \Psi_2(d,x,y,y_1,y_2) &\simeq& 
    \int_Z^1 dz\ \frac{ z^{3-\frac{3}{2}d} }{ [\ \ zD(y,y_1,y_2)\ ]^{5-2d} } 
  = D(y,y_1,y_2)^{2d-5} \ \int_Z^1 dz z^{\frac{d}{2}-2} 
                                                     \nonumber\\ 
 &=& \frac{2}{d-2} \ D(y,y_1,y_2)^{2d-5} \ , 
\label{eq:Psi2val} 
\end{eqnarray} 
where we have neglected the contribution from the lower integration 
limit $Z$, which is $Z^{\frac{d}{2}-1}$, as for $d$ just bigger than 2 
it vanishes with $Z$. \par 
The first term is slightly more delicate. To start with, 
as $0<z<Z\ll 1$ we can neglect $z$ with respect to 1 in $(1-z)^2\;x^2$ 
\[ \Psi_1(d,x,y,y_1,y_2) \simeq \int_0^Z dz\ \frac{z^{3-\frac{3}{2}d}} 
                         { [x^2+zD(y,y_1,y_2)]^{5-2d} } \ ; \] 
for reasons which will be apparent in a moment, we rewrite it as 
\[ \Psi_1(d,x,y,y_1,y_2) = \int_0^Z dz\ \frac{ z^{\frac{d}{2}-2} } 
            { \left[D(y,y_1,y_2)+\frac{x^2}{z}\right]^{5-2d} } \] 
and integrate by parts the factor $z^{\frac{d}{2}-2}$; the result is 
\begin{eqnarray} 
  \Psi_1(d,x,y,y_1,y_2) &=& \frac{2}{d-2} \left\{ \left. 
                      \frac{ z^{4-\frac{3}{2}d} } 
                   { [x^2+zD(y,y_1,y_2)]^{5-2d} } \right|_0^Z 
         + (2d-5)\ x^2 \int_0^Z dz\ \frac{ z^{3-\frac{3}{2}d} }
            { [x^2+zD(y,y_1,y_2)]^{6-2d} } \ . \right\} \nonumber\\ 
        &=& 2\frac{2d-5}{d-2}\ x^2 \int_0^Z dz\ 
          \frac{ z^{3-\frac{3}{2}d} } 
            { [x^2+zD(y,y_1,y_2)]^{6-2d} } \ , \nonumber 
\end{eqnarray} 
as the end-point contributions vanish for $d$ just bigger than $2$. We can 
now modify the integration interval for $ \Psi_1(d,x,y,y_1,y_2) $ into 
\[ \int_0^Z dz = \int_0^\infty dz - \int_Z^\infty dz \ , \] 
and write correspondingly 
\begin{equation} 
 \Psi_{1}(d,x,y,y_1,y_2) = \Psi_{1\infty}(d,x,y,y_1,y_2)  
                         - \Psi_{1Z}(d,x,y,y_1,y_2) \ , 
\label{eq:splitPsi1} 
\end{equation} 
where, thanks to the previous integration by parts, the two resulting 
$z$-integrals (from $0$ to $\infty$ for the first, from $Z$ to $\infty$ 
for the second) are now both convergent for $d$ just bigger than $2$. 
\par 
We start again from the second term, 
\[ \Psi_{1Z}(d,x,y,y_1,y_2) = 2\frac{2d-5}{d-2}\ x^2 \int_Z^\infty 
     dz\ \frac{ z^{3-\frac{3}{2}d} }{ [x^2+zD(y,y_1,y_2)]^{6-2d} } \ ; \] 
in the denominator we can neglect $x^2$, the resulting integral is 
trivial and the result can be written as 
\[ \Psi_{1Z}(d,x,y,y_1,y_2) = 2\frac{2d-5}{d-2}\ \frac{1}
                                { [D(y,y_1,y_2)]^{6-2d} } 
     \ \frac{2}{d-4}\ \left\{ \left(\frac{x^2}{z}\right) 
      \ z^{\frac{d}{2}-1} \right\}_Z^\infty \ , \] 
which vanishes with $Z$ (recall $x^2\ll Z$), so that 
\begin{equation} 
 \Psi_{1Z}(d,x,y,y_1,y_2) = 0 \ . 
\label{eq:Psi1Zval} 
\end{equation} 
In the first term of Eq.(\ref{eq:splitPsi1}), 
$ \Psi_{1\infty}(d,x,y,y_1,y_2), $ we substitute 
$z=t x^2/D(y,y_1,y_2)$, obtaining 
\[ \Psi_{1\infty}(d,x,y,y_1,y_2) = 
            2\frac{2d-5}{d-2}\ x^{(d-2)}\ [D(y,y_1,y_2)]^{\frac{3}{2}d-4} 
               \int_0^\infty dt\ \frac{ t^{3-\frac{3}{2}d} } 
                                     { (1+t)^{6-2d} } \ . \] 
By using the second representation of Euler's Beta function 
\begin{eqnarray} 
  B(\alpha,\beta) &=& 
   \frac{ \Gamma(\alpha)\Gamma(\beta) }{ \Gamma(\alpha+\beta) } \nonumber\\ 
 &=& \int_0^1 du\ u^{\alpha-1}\ (1-u)^{\beta-1} 
   = \int_0^\infty dt\ \frac{ t^{\alpha-1} }{ (1+t)^{\alpha+\beta} } \ , 
\label{eq:defBeta} 
\end{eqnarray} 
we finally obtain 
\begin{equation} 
   \Psi_{1\infty}(d,x,y,y_1,y_2) = 2\frac{2d-5}{d-2} 
         B\left(4-\frac{3}{2}d,2-\frac{1}{2}d\right) 
          \ x^{(d-2)}\ [D(y,y_1,y_2)]^{\frac{3}{2}d-4} \ . 
\label{eq:Psi1ooval} 
\end{equation} 
By collecting the results Eq.s(\ref{eq:Psi1ooval},\ref{eq:Psi1Zval}, 
\ref{eq:splitPsi1},\ref{eq:Psi2val},\ref{eq:defPsii}), one obtains 
the value of the function $\Psi(d,x,y,y_1,y_2)$ to be substituted in 
Eq.(\ref{eq:newPhi1}); recalling Eq.(\ref{eq:defD}) one sees that all 
the remaining integrations factorize and can be carried out in terms of 
Euler's Beta-functions Eq.(\ref{eq:defBeta}). \par 
It is clear that the final result for Eq.(\ref{eq:defPhi}) for 
$x\to0$ consists of two terms, the first constant (independent of $x$), 
the second proportional to $x^{d-2}$. In the notation of 
Eq.(\ref{eq:Phiexpexpl}) we have 
\begin{figure}[t] 
\begin{center} 
\includegraphics*[2cm,1cm][14cm,6cm]{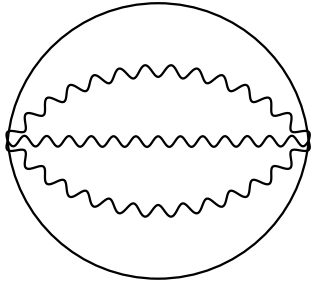} 
\caption{The 4-loop watermelon graph.} 
\end{center}
\label{fig:3} 
\end{figure} 
\begin{eqnarray} 
 A_0^{(1)} &=& -\ \frac{3d-11}{ 8(d-2)(d-3)(d-4)^3(2d-5)(2d-7)(3d-8)(3d-10) } 
                                             \nonumber\\ 
  &\times& \frac{ \Gamma(1-(d-4))\Gamma(1-2(d-4)) 
           \Gamma^2\left(1+\frac{1}{2}(d-4)\right) 
           \Gamma^2\left(1-\frac{3}{2}(d-4)\right) } 
         { \Gamma^4\left(1-\frac{1}{2}(d-4)\right) 
           \Gamma(1-3(d-4)) }                \nonumber\\ 
 A_0^{(2)} &=& -\ \frac{ 2\ (2d-7) } { 3(d-2)^2(d-3)(d-4)^4(3d-8)(3d-10) } 
                                             \nonumber\\ 
  &\times& \frac{ \Gamma\left(1+\frac{1}{2}(d-4)\right) 
                  \Gamma\left(1-\frac{3}{2}(d-4)\right) 
                  \Gamma^2(1-(d-4)) } 
                { \Gamma^2\left(1-\frac{1}{2}(d-4)\right) 
                  \Gamma(1-2(d-4)) } \ . 
\label{eq:Aval} 
\end{eqnarray} 
Let us observe that the term $A_0^{(1)}$ is the value of the vacuum graph
in Fig.3., in agreement with the result in Eq.(A.12) of \cite{vuoti} (up
to a different normalization).

\vspace*{1cm}\noindent{\Large {\bf Acknowledgments}} \\ 
\noindent 
We are grateful to J. Vermaseren for his kind assistance in the use 
of the algebra manipulating program {\tt FORM}~\cite{FORM}, by which 
all our calculations were carried out. 

 
\end{document}